\documentstyle[epsf,rotate,cite,12pt]{article}
\begin{document}
\title{SU(2) Gluon Propagator on a Coarse Anisotropic Lattice}
\author{{Ying Chen${}^{2}$\thanks{Email address:
cheny@hptc5.ihep.ac.cn},
         Bing He${}^{2}$, He Lin${}^{2}$, Ji-min Wu${}^{1,2}$}\\
{\small ${}^{1}$ CCAST (World Laboratory), P.O.Box 8730, Beijing
100080, P.R.China}\\
   {\small ${}^{2}$ Institute of High Energy Physics, Academia
Sinica, P.O.Box 918(4), Beijing 100039, P.R.China$\thanks{permanent
address}$}}
\maketitle
\begin{center}
\begin{minipage}{120mm}
\vskip 0.8in
\begin{center}{\bf Abstract}\end{center}
{We calculated the SU(2) gluon propagator in Landau gauge on an
anisotropic coarse lattice with the improved action. The standard and
the improved scheme are used to fix the gauge in this work. Even on the
coarse lattice the lattice gluon propagator can be well described 
by a function of the continuous momentum. The effect of
the improved gauge fixing scheme is found not to be apparent. Based on the 
Marenzoni's model,
the mass scale and the anomalous dimension are extracted and can be
reasonably extrapolated to the
continuum limit with the values $\alpha\sim 0.3$ and $M\sim 600MeV$. We
also extract the physical anisotropy $\xi$ from the gluon propagator due
to the explicit $\xi$ dependence of the gluon propagator.  
 } 
\end{minipage}

\end{center}

\newpage
\baselineskip 0.3in
\section*{I. Introduction}
The gluon propagator is well defined in the pure non-Abelian gauge theory
and is thought of one of the most basic quantities of QCD. Due to the
asymptotic freedom, the gluon propagator
at small distances (or large momentum) can be described
sufficiently by the perturbation theory in which the gluon is expected to
propagate as a massless particle. However, any asymptotic
gluon states and isolated quarks are absent, and this motivated the
confinement mechanism that gluons and quarks are only confined 
inside hadrons. This physical structure may imply that the gluon
propagator is quite different from those of stable particles. To understand 
the structure of the gluon propagator, especially its infrared behavior, 
a nonperturbative study of the gluon propagator is required. 
\par
Lattice simulation provides a powerful method to study the nonperturbative
properties of QCD. The first lattice simulations of the gluon propagator
were carried out by Mandula and Ogilvie\cite{mandula} and by Gupta {\sl
et al}\cite{gupta} in the late 1980s, in whose studies the gluon
propagator 
in Landau gauge was interpreted in terms of a massive particle
propagator. The simulations of Marenzoni {\sl et al}\cite{model} gave
the result that in terms of an anomalous dimension the propagator fell off
quite differently from that in
perturbation theory. The most recent
simulation carried out on a very large lattice by Leinweber {\sl et al}
gave the implication that the gluon propagator should be
interpreted in terms of neither a massless particle nor a free massive
one\cite{leinweber}. However, even though there has been substantial
progress in the lattice simulation of gluon propagator in the last
decade, a definitive picture of the gluon propagator has not been obtained
yet. For reference see the overview of Mandula\cite{mandula1}.  
\par
The development of the improved actions is an important progress in
lattice QCD in last decade. If the lattice action is properly improved,
the costs of the Monte Carlo simulation will be greatly reduced and the
numerical study can be performed on much coarser lattice system.
Ma\cite{majp} simulated the SU(3) gluon propagator in Landau gauge on a
coarse lattice using tadpole improved Symanzik's action with substantially
reduced lattice artifact. In his work the propagator was fitted using 
Marenzoni's model and gave the result that the mass scale turns out good
scaling behavior and the anomalous dimension can be well extrapolated to
the continuum limit $a\rightarrow 0$. 
\par
Since the anisotropic lattice has a lot of advantages in the Monte carlo
simulation of lattice QCD on coarse lattice, it is meaningful to study the
behavior of
gluon propagator on anisotropic lattice and the effect of this
temporal-spatial asymmetric lattice on the gluon propagator because the
lattice definition of gluon propagtor is explicitly dependent on the
anisotropy ratio $\xi$. Another goal of this work is to explore the
efficiency of the improved guage fixing scheme introduced by Bonnet {\sl
et al}\cite{bonnet}. This work is organized as follows. In
section II we describe the lattice formulations used in this work. Section
III is devoted to the simulation details where we give the mass scales and
the anomalous dimensions extracted from the lattice propagator as well as
their lattice spacing dependence. In Section IV we discuss the effects of
the asymmetric lattice on the gloun propagator and extract the anisotropy
ratio for various lattice spacings. Section V is the summary.

\section*{II. Lattice Formulations}
We use the naive steepest descent method to fix the
Landau guage\cite{davies} by maximising the usual Landau gauge fixing
functional. There are many choices of the definition of the gauge fixing
functional. The frequently used version (called standard one in this work)
is 
\begin{equation}
\label{fun1}
{\cal F}_1^G[\{U\}]=\sum\limits_{x,
\mu}\frac{1}{2}Tr\left\{U_{\mu}^G(x)+U_{\mu}^G(x)^{\dagger}\right\},
\end{equation}
where
\begin{equation}
U_{\mu}^G(x)\equiv G(x)U_{\mu}(x)G(x+\hat{\mu})^{\dagger}
\end{equation}
and 
\begin{equation}
G(x)=exp\left\{-i\sum\limits_a\omega^a(x)T^a\right\}.
\end{equation}
Taking the extremum condition  
\begin{equation}
\frac{\delta{\cal
F}_1^G}{\delta\omega^a(x)}\propto\sum\limits_{\mu}
\left[U_{\mu}^G(x-\hat{\mu})-U_{\mu}^G-
\left(U_{\mu}^G(x-\hat{\mu})-U_{\mu}^G\right)^{\dagger}\right]
\equiv \Delta_1(x)=0
\end{equation}
and expanding the gauge links, $U_{\mu}(x)={\cal 
P}exp\left\{iag\int\limits_0^1 dt A_{\mu}(x+\hat{\mu}at)\right\}$, in the
small lattice spacing, one obtains the Landau gauge condition on
the lattice as 
\begin{equation}
\sum\limits_{\mu}\partial_{\mu}A_{\mu}(x)=\sum\limits_{\mu}\left\{-\frac{a^2}
{12}\partial_{\mu}^3 A_{\mu}(x)+O(a^4)\right\}.
\end{equation}
The lattice artifact to the gauge condition is of
order $O(a^2)$. An alternative gauge fixing functional is introduced by
Bonnet {\sl et al}\cite{bonnet} as follows,
\begin{equation}
{\cal F}_2^G[\{U\}]=\sum\limits_{x,\mu}\frac{1}{2}Tr\left\{
U_{\mu}^G(x)U_{\mu}^G(x+\hat{\mu})+h.c.\right\}.
\end{equation}
Similarly the equation
\begin{eqnarray}
\frac{\delta{\cal F}_2^G}{\delta\omega^a(x)}&\propto&
\sum\limits_{\mu}Tr\left[U_{\mu}^G(x-2\hat{\mu})U_{\mu}^G(x-\hat{\mu}) 
-U_{\mu}^G(x)U_{\mu}^G(x+\hat{\mu})-h.c.\right]\\\nonumber
&\equiv& \Delta_2=0
\end{eqnarray}
gives another realization of lattice Landau gauge,
\begin{equation} 
\sum\limits_{\mu}\partial_{\mu}A_{\mu}(x)=\sum\limits_{\mu}\left\{-\frac{a^2}
{3}\partial_{\mu}^3 A_{\mu}(x)+O(a^4)\right\}.
\end{equation}
The $O(a^2)$ artifact can be removed by combining 
the two types of the gauge fixing functional to an improved
one\cite{bonnet}, ${\cal
F}_{Imp}^G\equiv \frac{4}{3}{\cal F}_1^G-\frac{1}{12u_{\mu}}{\cal
F}_2^G$,
where $u_{\mu}$ refers to the tadpole improved parameter of the gauge
links in $\mu$ direction. The improved functional gives the following
new function
\begin{equation}
\Delta_{Imp}(x)=\frac{4}{3}\Delta_1(x)-\frac{1}{3u_{\mu}}\Delta_2(x).
\end{equation}
The steepest descents approach is realized by defining the local gauge
transformation as
\begin{eqnarray}
G_i(x)&=&exp\left\{\frac{\alpha}{2}\Delta_i(x)\right\}\\\nonumber
    &\approx&1+\frac{\alpha}{2}\Delta_i(x),
\end{eqnarray}
where $\alpha$ is a tuneable step-size prameter. The lattice  
Landau gauge is fixed by decreasing the quantity,  
\begin{equation}
\theta_i=\frac{1}{2V}\sum\limits_{x}Tr\left\{\Delta_i(x)
\Delta_i(x)^{\dagger}\right\},
\end{equation}
to a sufficiently small number. Here the suffix $i$ stands for 1, 2 and
Imp. 
\par
In this work, the simulations are performed on an $8^3\times 24$
anisotropic lattice by using tadpole-improved Symanzik's
action\cite{morningstar,lepage}. The lattice spacing in the temporal
direction is smaller than that in the spatial direction with the
bare anisotropy 
ratio $\xi_0$. The action we use is given as
\begin{eqnarray}
\label{action}
S_1 &=& \beta \sum\limits_{x,s>s'}\left[\frac{5}{3}\frac{P_{ss'}}{\xi_0
u_s^4}-\frac{1}{12}\frac{R_{ss'}}{\xi_0 u_s^6}
-\frac{1}{12}\frac{R_{s's}}{\xi_0 u_s^6}\right]\\\nonumber
&+&\beta \sum\limits_{x,s}\left[\frac{4}{3}\frac{P_{st}\xi_0}{u_s^2 u_t^2}
-\frac{1}{12}\frac{R_{st} \xi_0}{u_s^4 u_t^2}\right],
\end{eqnarray}  
where $\beta=\frac{4}{g^2}$, $g$ is the gauge coupling. $P_{\mu
\nu}$ is the plaquette operator and  $R_{\mu \nu}$  $2\times 1$
rectangular operator, $u_s$ and $u_t$ are tadpole improvement  
parameters. $u_s$ is defined by $u_s=<(1/2)TrW_{sp}>^{1/4}$, while $u_t$
is set to be 1 because of $u_t=1-O(\xi_0^{-2})$ ($\xi_0$ is much bigger
than one).
\par 
The continuum gluon propagator $D_{\mu\nu}^{ab}$ in Landau gauge is
characterized by a single function $D(q^2)$ as
\begin{equation}
D_{\mu\nu}^{ab}(q)=\delta_{ab}\left(\delta_{\mu\nu}-
\frac{q_{\mu}q_{\nu}}{q^2}\right)D(q^2),
\end{equation}
where $a$ and $b$ are the color index. The structure of $D(q^2)$ is of
physical interest. In the large momentum range $D(q^2)$ takes its
perturbative form and is proportional to $1/q^2$, while its infrared
behavior may be more complicated and should be reconciled with the 
confinement. The lattice simulation of the gluon propagator is to
determine its structure nonperturbatively. The lattice version of
$D(q^2)$ is denoted as $D_L(q)$, which can be constructed in terms of
the Fourier transfromed vector potential $A_\mu(q)$. The vector
potential of SU(2) gauge field is defined from the gauge links
\begin{eqnarray}
A_{\mu}(x+a_{\mu}\hat{\mu}/2)&=&\frac{1}{2iga_{\mu}}(U_{\mu}(x)-
U_{\mu}^{\dagger}(x)-\frac{1}{2}Tr(U_{\mu}(x)-U_{\mu}^{\dagger}(x)))\\\nonumber
&\equiv&\frac{1}{a_\mu}\hat{A}_{\mu}(x+a_{\mu}\hat{\mu}/2).
\end{eqnarray}
where $a_{\mu}$ is the lattice spacing in $\mu$ direction, and 
its  Fourier transformation is defined as
\begin{equation}
A_{\mu}(q)=\left(\prod\limits_{\mu}a_\mu\right)
\sum\limits_{x}e^{-iq\cdot(x+a_{\mu}\hat{\mu}/2)}
A_{\mu}(x+a_{\mu}\hat{\mu}/2)\equiv \frac{1}{a_\mu}\hat{A}_{\mu}(q),
\end{equation}
where $q$ is the momentum
\begin{equation}
\label{momentum}
q_{\mu}=\frac{2\pi}{L_{\mu}
a_{\mu}}n_{\mu},~~~~n_{\mu}=0,1,\ldots,L_{\mu}-1.
\end{equation}  
Here the notation $\hat{A}$ is adopted for the convenience of the
future discussion. The function $D_L(q)$ is defiend as 
\begin{equation}
D_L(q)=\frac{2}{9\prod\limits_\mu(L_\mu a_\mu)}\sum\limits_{\mu}\langle
0|Tr(A_{\mu}(q)A_{\mu}^{\dagger}(q)|0\rangle,
\end{equation}
which approaches to the function $D(q^2)$ in the limit of
$a_\mu\rightarrow 0$.
\section*{III. Lattice Simulation}
We used the action defined in Eqn.(\ref{action}) to generate gluonic
configurations at $\beta=1.0,~1.1,~1.2$ on an $8^3\times 24$ anisotropic
lattice. The
samples of configurations and the values of $u_s$ and $a_s$\cite{zhang} 
for various $\beta$ are listed in Table 1. The configurations are
separated by 50 Monte carlo sweeps, each of which is composed of 4
heatbath and a microcanonical iterations. We performed the standard gauge
fixing and the improved gauge fixing for each configuration respectively
and stopped the gauge transfromation at $\theta_i<10^{-8}$. The Landau
gauge is surely reached for each configuration, for it was observed that
for any $q=({\bf q},0)$, $trA_4(q)A_4^{\dagger}(q)$ are of order $10^{-6}$
and are about five order smaller than other components.
\par
It is expected that the function $D_L(q)$ measured on the lattice
approximately depend only on $q^2=\sum\limits_{\mu}q_{\mu
}^2$ for the small enough $a_{\mu}q_{\mu}$. Previous
simulations, which were performed on large lattice with the Wilson action,
gave the results that this is only true for a small range $a^2q^2<1$. By
applying the improved action we find that the function $D_L(q)$ from 
our simulation behaves well as a function of $q^2$ in a much larger 
momentum range $0<a^2q^2<9$ which corresponds to almost the full first
Brillouin zone (see Figure 1a, 1b, 1c,
the data points are
the averages of $D_L(q)$ with the same $q^2$). This range corresponds
to almost the full first Brillouin zone. From now on we denote
$D_L(q)$ as $D_L(q^2)$.
\par
We use the model\cite{model} to fit the data points of $D_L(q^2)$,
\begin{equation}
D_L(q^2)=\frac{Z}{(M^2)^{1+\alpha}+(q^2)^{1+\alpha}},
\end{equation}
where $M$ is a dimensional parameter, $\alpha$ the anomalous
dimension and $Z$ a parameter related to $g$ and the renormalization
constant of wave-function. We perform the fitting in two ways which use
the data sets including (fit 1) and excluding (fit 2) the zero momentum
point respectively, and the results are listed in Table 2 and Table 3. The
quality of both fits are acceptable due to the very small $\chi^2/d.o.f$.
It is found that the effect of the improvement of gauge fixing
is not apparent, for the fitting results of $D_L(q^2)$ through the
standard and the improved gauge fixing procedures are almost the same. The
mass scale $M$ in physical unit is listed in Table 4. In fit 2, the
mass scale is approximately $600 MeV$ within the errors and insensitive to
the lattice spacing, while in fit 1 the differences of $M$ for various
lattice spacing are significant.  
\par
The anomalous dimension $\alpha$ is extracted from both the fits. In
fit 1 the value of $\alpha$ varies with $\beta$ and the $a_s$ dependence
of $\alpha$ can be well described as 
\begin{equation}
\alpha(a_s)=\alpha_0+\alpha_1 a_s^4,
\end{equation}
with
$$ \alpha_0=0.309(10)~~~\alpha_1=-19.9(1.2)~~~\chi^2=0.097.$$

In fit 2, the values of $\alpha$ are averaged to be
$\bar{\alpha}=0.282(15)$ and less related to the lattice spacing. 
\par
Even though the small discrepancy, both of the fits give the non-zero
anomalous dimension $\alpha\sim 0.3$.
This value is compatible with the
result of Ma's simulation of SU(3) gluon
propagator\cite{majp}'($\alpha=0.285(20)$) but smaller than the results of
Marenzoni\cite{model} ($\alpha \sim 0.4-0.6$). 
\section*{IV. The Renormalized Anisotropy Ratio}
Since the MC simulation is performed on an anisotropic lattice, there
naturally arises a quesiton: what is the effect of this asymmetry
of the temporal and the spatial direction on the physical observables and
how differently the physical obsevables are affected the renormalized
anisotropy ratio from the bare parameter $\xi_0$ ( here the bare parameter
refers to the parameter in the action which justifies that the
lattcie action would take the same form as that of the continuum action at
the classical level when the continuum limit $a\rightarrow 0$ is
approached, while the renormalized one, $\xi=\xi(\xi_0, \beta)$ is defined
by the ratio $a_s/a_t$ 
and is a funciton of the bare one and the coupling constant owing to the 
renormalization effects.)
This question
has been explored by Klassen {\sl et al.} who extracted $\xi$ from
the quark static potential\cite{klassen}. We propose here a new scheme to
extract the renormalized anisotropy ratio from the gluon propagator
measured on the anisotropic lattice.
\par
The lattice version of the funciton $D_L(q^2)$ and the continuous
momentum are  explicitly related to
$\xi$ as follows
\begin{eqnarray}
D_L(q)&=&\frac{2}{9\prod\limits_{\mu}(L_\mu a_\mu}\sum\limits_{\mu}\langle
0|Tr(A_{\mu}(q)A_{\mu}^{\dagger}(q)|0\rangle\\\nonumber
&\propto&\left\langle
0\left|Tr\left(\sum\limits_{i=1}^{3}\hat{A}_i(q)\hat{A}_i^{\dagger}+\xi^2
\hat{A}_4(q)\hat{A}_4^{\dagger}\right)\right|0 \right\rangle\\\nonumber
&\equiv& \hat{D}_L(q,\xi).
\end{eqnarray}
and 
\begin{equation}
q^2=\left(\frac{2\pi}{L_s a_s}\right)^2\left(\sum\limits_i
n_i^2+\frac{\xi^2}{9}n_4^2\right).
\end{equation}

In practice, the renormalized anisotropy $\xi$ is unknown at the
beginning of the calculation and the measured quantity is
$D_L(q,\xi_0=3)$. For convenience we introduce a notation
$\hat{q}^2=\sum\limits_{\mu}n_{\mu}^2$ where $n_\mu$ is defined
in the Eqn.(\ref{momentum}). If the function $D_L(q)$ is a
function of $q^2$ and $\xi/\xi_0=1$, the rotational invariance requires
that $D_L(q)$ take the same value at all the momentum with the same
$\hat{q}^2$. From the analysis of our results we found that this is only
correct 
for the momenuta with the same $\hat{q}^2$ and the same $n_4$. For
those momenta with the same $\hat{q}^2$ but different $n_4$, the
difference of $D_L(q)$ is significant. This discrepancy can be
attributed to the fact that $\xi/\xi_0\neq 1$, as can be found in the 
following. 
\par 
With the conjecture that $\epsilon/\xi_0$ is small, only the lowest order
term of $\epsilon=\xi_0-\xi$ is relevant in the discussion,
$D_L(q,\xi_0=3)$ is
related to the expected $D_L(q,\xi)$ as follows,
\begin{equation}
\hat{D}_L(q,\xi)=\hat{D}_L(q,3-\epsilon)=\hat{D}_L(q,3)-6\epsilon
\langle 0|Tr\hat{A}_4(q)\hat{A}_4^{\dagger}|0\rangle+O(\epsilon^2).
\end{equation}
For simplicity we define
\begin{equation}
\alpha(q)=\langle 0|Tr A_4(q)A_4(q)^{\dagger}|0\rangle.
\end{equation}
Another reason for the discrepancy is the $\xi$ dependence of the
continuous momentum also introduces a small shift $\Delta q^2$ of momentum
square
between two momenta $q$ and $q'$ with the same $\hat{q}^2$. With both
the two factors in consideration and accepting the fact that
$D_L(q,\xi)$ is well described by a function of $q^2$, this discrepancy,
described by $\eta=\frac{\hat{D}_L(q',3)}{\hat{D}_L(q,3)}$, can be
re-expressed in the lowest order of $\epsilon=\xi$,
\begin{eqnarray}
\label{epsilon}
\eta&=&\frac{\hat{D}_L(q',3)}
{\hat{D}_L(q,3)}=\frac{\hat{D}_L(q',\xi)+6\epsilon\alpha(q')}
{\hat{D}_L(q,3)+6\epsilon\alpha(q)}\\\nonumber
&=&\frac{\hat{D}_L(q',\xi)}{\hat{D}_L(q,\xi)}
\left[1+6\epsilon\left(\frac{\alpha(q')}{\hat{D}_L(q',\xi)}-
\frac{\alpha(q)}{\hat{D}_L(q,\xi)}\right)\right]\\\nonumber
&=&\left(1-\Delta q^2\frac{d}{dq^2}\ln
\hat{D}_L(q^2,\xi)|_{q^2=q'^2}\right)\left[1+\frac{6\epsilon}{\hat{D}_L(q,\xi)}
\left(\frac{\hat{D}_L(q,\xi)}{\hat{D}_L(q',\xi)}\alpha(q')
-\alpha(q)\right)\right].
\end{eqnarray}
Since only the lowest order of $\epsilon$ is considered, after replacing
approximately the ratio $\frac{\hat{D}_L(q',\xi)}{\hat{D}_L(q,\xi)}$ with
$1/\eta$ and rearranging  the Eqn.(\ref{epsilon}), the quantity $\epsilon$
can be expressed by the quantities which can be measured directly from the
lattice simulation (see the following equation),
\begin{equation}
\epsilon(q^2)=\frac{1-\eta}{\frac{2}{3} \delta(q^2) \frac{d}{dq^2} \ln
\hat{D}_L(q,\xi)+\frac{6}{\hat{D}_L(q,\xi)}\left(\alpha(q)-\frac{1}{\eta}
\alpha(q')\right)}.
\end{equation}
Here the momenta $q$ and $q'$ satisfy $\hat{q}^2=\hat{(q')}^2$ with
$\delta(q^2)=\frac{2}{3}\epsilon ((q')_4^2-q_4^2)$. The results of
$\epsilon(q^2)$ for $\beta=1.0, 1.1, 1.2$ are listed in Table \ref{E}.
Theoretically, $\xi/\xi_0$ should be independent of $q^2$. From the
Table it is found that the measured $\xi/\xi_0$ are insensitive to $q^2$
for most momenta. However, a few irregular values  
such as $q^2(\delta(q^2)=6(1)$
for $\beta=1.0$, $q^2(\delta(q^2)=4(1)$ for $\beta=1.1$ and
$q^2(\delta(q^2)=4(1)$ for $\beta=1.2$, are a bit larger than the others.
We attribute it to the reason that the rotational invariance is
violated to some extent so that the value of $D_L(q)$ with momenta
directed along one of the four axes is different from those with
momenta directed off-axis. Averaging the regular values we obtain
$$\xi/\xi_0=1.23~~~for~~~\beta=1.0$$
$$\xi/\xi_0=1.15~~~for~~~\beta=1.1$$
$$\xi/\xi_0=1.16~~~for~~~\beta=1.2.$$
These are raw results at the lowest order of $\epsilon$ and we do not
plan to extract the $\beta$ dependence of $\xi/\xi_0$ from only three
data points. What we can conclude at present is that the
renormalization effects for the anisotropy ratio is significant and our
results is consistent with those from the study of static quark potential
by Klassen {\sl et al}\cite{klassen}. Morningstar {\sl et al}\cite{morn1} 
have simulated the SU(3) glueball mass on the anisotropic lattice with the
same action as that we use and reported that $\xi/\xi_0$ extracted from
the static quark potential is apporximately one within few percents
discrepancy. Work is underway to explain the reason why the anisotropy
ratios extraceted from different quantities do not comply with each other.

\section*{V. Summary}
We studied the SU(2) gluon propagator in Landau gauge on an $8^3 \times
24$ anisotropic lattice by using the tadpole improved Symanzik's action.
Although the improvement scheme is also used to fix the gauge, we found
that its effect is not significant, while the lattice
artifact is significantly suppressed by using the improved action.
The function $D_L(q)$ in the propagator can be well described as a
function of continuous momentum square $q^2$ even in the full first
Brillouin zone. We used the model suggested
by Marenzoni {\sl et al} to fit the function $D_L(q)$ and extracted the
mass scale and the anomalous dimension therein. It is found that the value
of the mass scale is approximately independent of the lattice spacing and
the anomalous dimension can be well extrapolated to the value about 0.3 in
the continuum limit.
\par
The effects of the asymmetric lattice on the gluon propagator are also
explored in this work. Based on the fact that the lattice definition
of the gluon propagator is related to the renormalized anisotropy ratio
$\xi=a_s/a_t$, we deduced a formula to measure the $\xi$ nonperturbatively
at various momentum. Even in the lowest order of $\epsilon=\xi_0-\xi$,
the measured values of $\xi$ are insensitive to the momentum and are
consistent with the results extracted from the static quark potential.

\section*{Acknowledgement}
This work is supported by the Natural Science Foundation of China under
the Grant No. 19677205 and No. 19991487, and by the National Research
Center for Intelligent Computing System under the contract No. 99128.

\section*{Tables and Figures}
\begin{table}[hbp]
\begin{center}
\begin{tabular}{|c|c|c|c|}\hline
$\beta$       &$u_s$      & $a_s$(fm)    &      No. of Configs.\\\hline
 1.0	      & 0.8215     & 0.277	  &      200\\\hline
 1.1          & 0.8365     & 0.233        &      200\\\hline
 1.2          & 0.8593     & 0.183        &      200\\\hline
\end{tabular}
\caption[]{The parameters used in the simulations of this work.}
\end{center}
\end{table}

\begin{table}[hbp]
\begin{center}
\begin{tabular}{|c|c|c|c|c|c|c|}\hline
\multicolumn{1}{|c|}{}&\multicolumn{3}{|c|}{Standard fixing}
&\multicolumn{3}{|c|}{Improved fixing}\\\hline
$\beta$ & $a_s^2 M^2$ & $\alpha$ &  $\chi^2/dof$& $a_s^2 M^2$
& $\alpha$ &$\chi^2/dof$\\\hline
1.0 & 0.552(24) & 0.191(15) &0.17 & 0.547(24) & 0.203(15) &0.17\\\hline
1.1 & 0.357(14) & 0.254(13) &0.17 & 0.360(14) & 0.268(13) &0.15\\\hline
1.2 & 0.311(12) & 0.285(13) &0.13 & 0.313(11) & 0.298(13)& 0.12\\\hline
\end{tabular}
\caption[]{The results of the fit 1 which includes the zero-momentum data
point. $\alpha$ and $a_s^2 M^2$ in the propagator model
are extracted at various value of $\beta$. The data illustrate that the
effect of the improved gauge fixing scheme is not apparent.}
\end{center}
\end{table}

\begin{table}[hbp]
\begin{center}
\begin{tabular}{|c|c|c|c|c|c|c|}\hline
\multicolumn{1}{|c|}{}&\multicolumn{3}{|c|}{Standard fixing}
&\multicolumn{3}{|c|}{Improved fixing}\\\hline
$\beta$    & $a_s^2 M^2$ & $\alpha$ & $\chi^2/dof$ & $a_s^2
M^2$ &  $\alpha$ &$\chi^2/dof$\\\hline
1.0 & 0.808(95) & 0.273(33) &0.11 &0.807(88) &0.288(32) & 0.11\\\hline
1.1 & 0.454(70) & 0.289(28) &0.18 &0.439(69) &0.297(28) & 0.17\\\hline
1.2 & 0.309(66) & 0.284(27) &0.15 &0.317(66) &0.299(27) & 0.13\\\hline
\end{tabular}
\caption[]{The results of the fit 1 which excludes the zero-momentum data
point. $\alpha$ and $a_s^2 M^2$ in the propagator model
are extracted at various value of $\beta$. The data illustrate that the
effect of the improved gauge fixing scheme is not apparent.} 
\end{center}  
\end{table}

\begin{table}[hbp]
\begin{center}
\begin{tabular}{|c|c|c|c|}\hline
$\beta$	& 1.0		& 1.1		&1.2	\\\hline
M(MeV)(Fit 1)	& 529(12)	& 506(10)	&601(12)\\\hline
M(MeV)(Fit 2) 	& 640(37)	& 571(44)	&599(64)\\\hline
\end{tabular}
\caption[]{Tha mass scales in physical unit. In fit 2, within the
range of errors, the mass scale takes approximately the same value and is 
independent of the lattice spacing.}
\end{center}
\end{table}

\begin{table}[hpb]
\label{E}
\begin{center}
\begin{tabular}{|c|c|c|c|c|c|c|}\hline
\multicolumn{1}{|c|}{}&
\multicolumn{2}{|c|}{$\beta=1.0$}&
\multicolumn{2}{|c|}{$\beta=1.1$}&
\multicolumn{2}{|c|}{$\beta=1.2$}\\\hline

$q^2(\delta(q_4^2))$   & 
$\epsilon$   &$\xi/\xi_0 $   &
$\epsilon$   &$\xi/\xi_0 $   &
$\epsilon$   &$\xi/\xi_0 $  \\\hline
1(1) & -1.04 & 1.35  &  -0.31 &  1.10 &  -0.40 &  1.13  \\\hline
2(1) & -0.69 & 1.23  &  -0.52 &  1.17 &  -0.50 &  1.17  \\\hline
3(1) & -0.79 & 1.26  &  -0.32 &  1.11 &  -0.27 &  1.09  \\\hline
4(1) & -0.66 & 1.22  &  -1.14 &  1.38 &  -1.09 &  1.36  \\\hline
4(4) & -0.54 & 1.18  &  -0.40 &  1.13 &  -0.39 &  1.13  \\\hline
5(1) & -0.75 & 1.25  &  -0.40 &  1.13 &  -0.50 &  1.17  \\\hline
5(4) & -0.63 & 1.21  &  -0.41 &  1.14 &  -0.43 &  1.14  \\\hline
6(1) & -1.09 & 1.36  &  -0.53 &  1.18 &  -0.58 &  1.19  \\\hline
6(4) & -0.68 & 1.23  &  -0.50 &  1.17 &  -0.49 &  1.16  \\\hline
7(1) & -0.85 & 1.28  &  -0.44 &  1.15 &  -0.62 &  1.21  \\\hline
8(4) & -0.71 & 1.24  &  -0.63 &  1.21 &  -0.60 &  1.20  \\\hline
\end{tabular}
\caption[]{The difference of the renormalized anisotropy ratio and the
bare one $\epsilon=\xi_0-\xi$ is measured at various $\hat{q}^2$. The
ratio $\xi/\xi_0$ is insensitive to the momentum except for a few
irregular values. These irregular values may be the result of the
significant violation of the rotational invariant.}
\end{center} 
\end{table}
\begin{figure}
\label{propa}
\epsfysize=2.5in
\hspace{3.5cm}
\epsffile{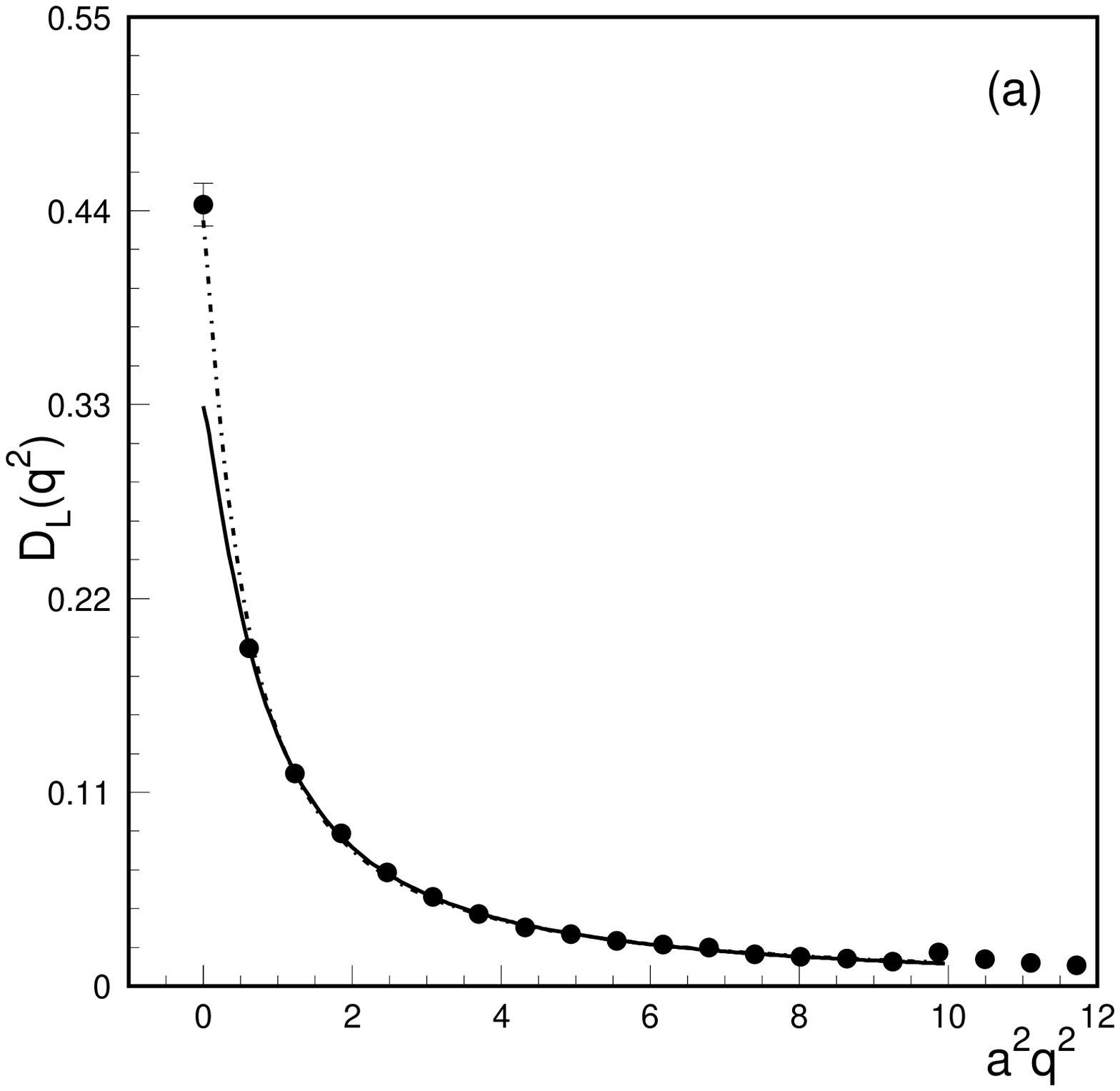}
\epsfysize=2.5in
\hspace{3.5cm}
\epsffile{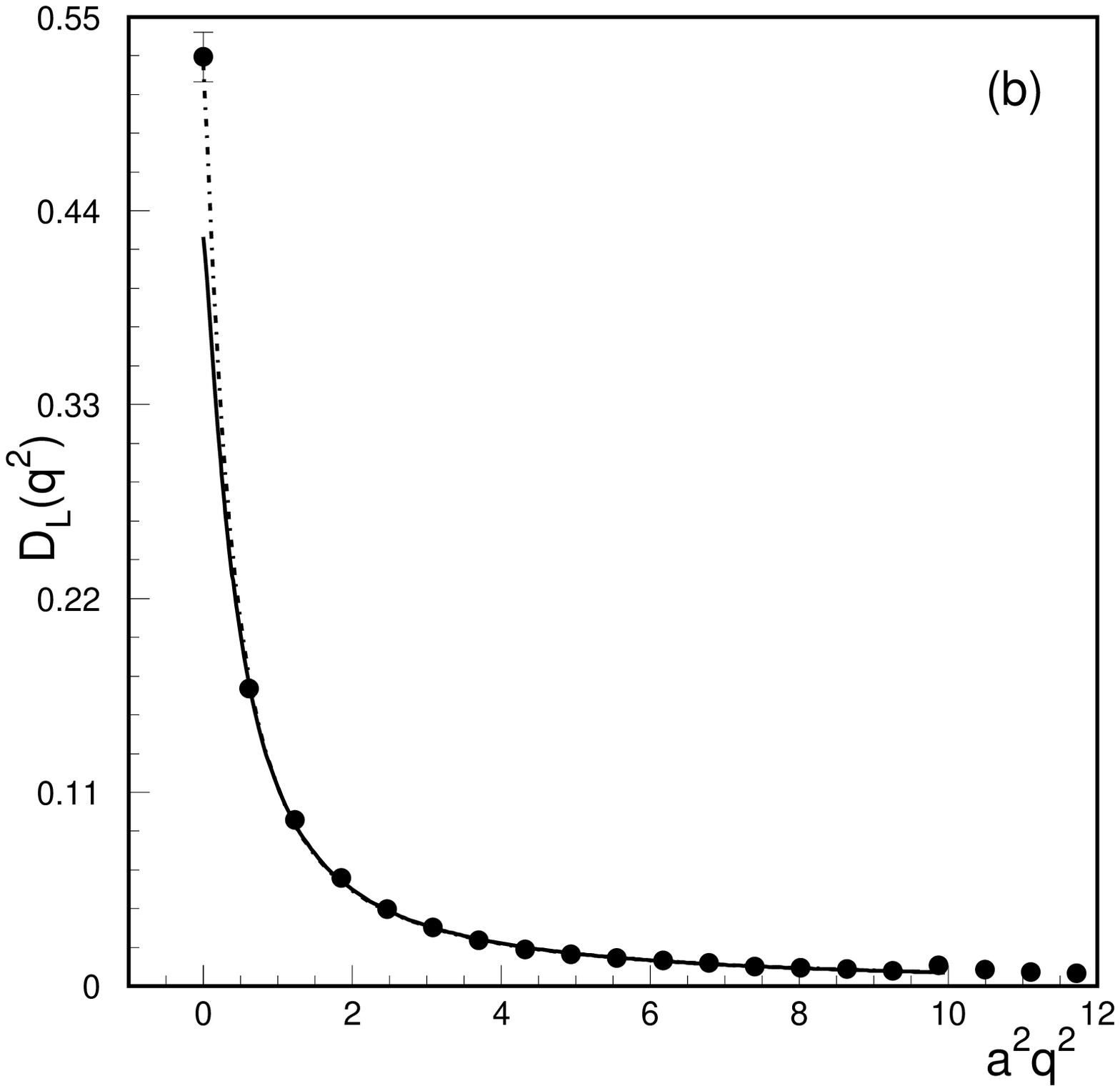}
\epsfysize=2.5in
\hspace{3.5cm}
\epsffile{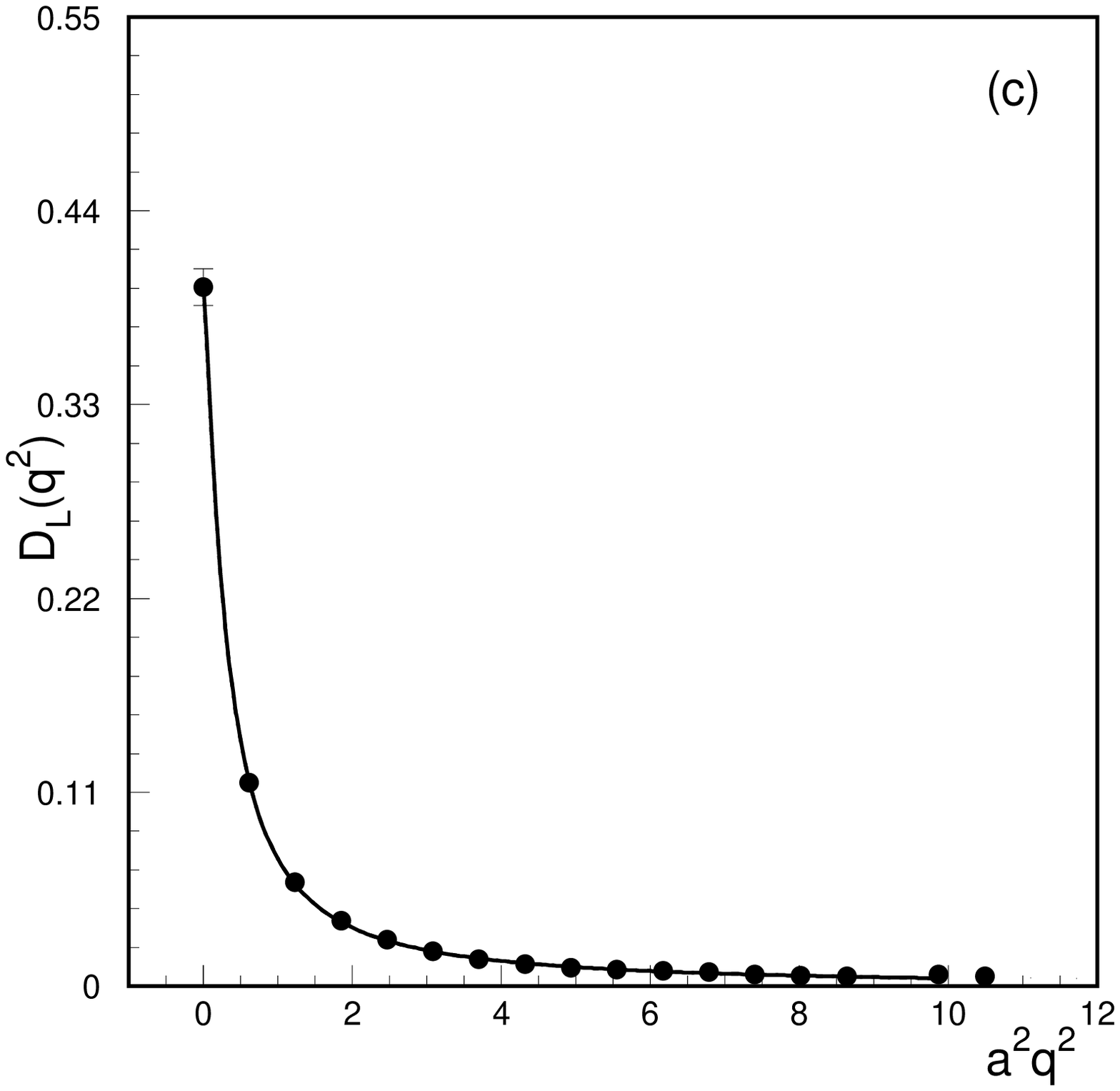}
\caption[]{The lattice gluon propagator $D_L(q)$ versus the
dimensionless continuous momentum $a^2 q^2$ measured at various
$\beta$. The filled circles are the data points from the simulation, the
dash line shows the result of fit 1 and the full line the
results of fit 2. a) $\beta=1.0$; b) $\beta=1.1$; c) $\beta=1.2$.
}
\end{figure}
\begin{figure}
\label{alpha}
\epsfysize=3in
\hspace{3.5cm}
\epsffile{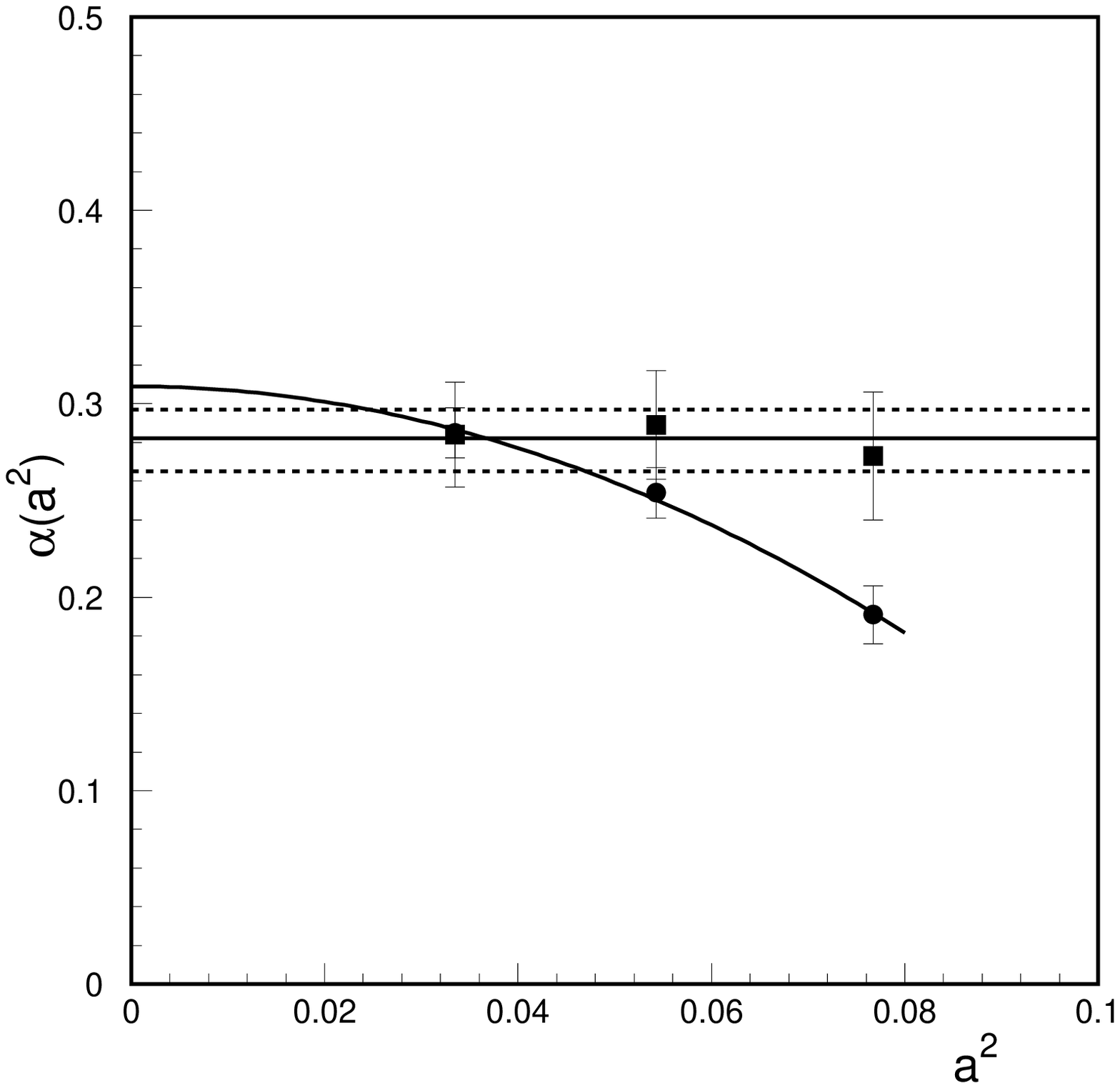}
\caption[]{The lattice spacing dependence of the anomalous
dimension
$\alpha$ is illustrate in the plot. $\alpha$ through fit 1 (filled
circles) can be extrapolated to the continuum limit by the model
$\alpha(a)=0.309-19.9 a^4$ with the $\chi^2=0.097$ while $\alpha$ through
fit 2 (filled square) is approximately independent of the lattice spacing
within the errors.}
\end{figure}

\begin{thebibliography}{s2}
\bibitem{mandula}J. E. Mandula and M. Ogilvie, Phys. Lett. B185(1987)127.
\bibitem{gupta}R. Gupta, G. Guralnik, G. Kilcup, A. Patel, S. Sharpe, T.
Warnock, Phys. Rev. D36(1987)2813.
\bibitem{model}P. Marenzoni {\sl et al.}, Phys. Lett. B318 (1993)511; P.
Marenzoni, G. Martinelli, and N. Stella, Nucl. Phys. B455(1995) 339.
\bibitem{leinweber}Leinweber, Skullerud, Williams, and Parrinello, Phys.
Rev. D (Rapid Commun.)58 (1998) 031501-1, hep-lat/9811027.
\bibitem{mandula1}J. E. Mandula, Phys. Rep. 315 (1999) 273.
\bibitem{majp}J. P. Ma, Mod. Phys. Lett. A15 (2000)229, hep-lat/9903009. 
\bibitem{bonnet}F. D. R. Bonnet, P. U. Bowman {\sl et al}, Austral. J.
Phys. 52 (1999)939, hep-lat9905006.
\bibitem{davies}C. T. H. Davies {\sl et al}, Phys. Rev. D37(1988)1581.
\bibitem{morningstar}C. J. Morningstar and M.
Peardon, Nucl. Phys. B(Proc. Suppl.)47(1996)258.
\bibitem{lepage}G. P. Lepage, P. B. Mackenzie, Phys. Rev.
D48(1993)2250.
\bibitem{zhang}J. B. Zhang, M. Jin and D. R. Ji, Chin. Phys. Lett.
15(1998)865; Y. Chen, B. He, H. Lin and J. Wu, Commun. Theor.
Phys.(in press).
\bibitem{klassen}T. R. Klassen, Nucl. Phys. B533(1998) 557.
\bibitem{morn1}C. Morningstar, M. Peardon, Phys. Rev. D60 (1999) 034509,
hep-lat9901004. 
\end{thebibliography}
\end{document}